\documentclass{article} 
\usepackage{graphicx}
\usepackage{amsmath}
\usepackage{amsfonts}
 \usepackage{amssymb}
\usepackage{ulem} 
\usepackage{xcolor}
\usepackage[square,numbers]{natbib}
\newtheorem{theorem}{Proposition}
\begin{document}
\begin{center}
\begin{LARGE}
\textbf{A minimal model coupling communicable and non-communicable diseases}
\end{LARGE}
\end{center}


\begin{center}
M. Marv\'a$^*$,\, 
        E. Venturino$^{**}$,\, M.C. Vera$^*$
\end{center}
\begin{center}
$^*$ Universidad de Alcal\'a, Departamento de F\'isica y Matem\'aticas. \\ Member of the research group Nonlinear Dynamics and Complex Systems.\\  Alcal\'a de Henares, Spain.\\

$^{**}$Dipartimento di Matematica "Giuseppe Peano".\\ Member of the INdAM research group GNCS.\\  Univertit\'a di Torino, Torino, Italy.\\
\end{center}

\begin{abstract} 
This work presents a model combining the simplest communicable and non-communicable disease models. The latter is, by far, the leadingn cause of sickness and death in the World, and introduces basal heterogeneity in populations where communicable diseases evolve. The model can be interpreted as a risk-structured model, another way of accounting for population heterogeneity. 

Our results show that considering the non-communicable disease (in the end, heterogeneous populations) allows the communicable disease to become endemic even if the basic reproduction number is less than $1$. This feature is known as subcritical bifurcation. Furthermore, ignoring the non-communicable disease dynamics results in overestimating the reproduction number and, thus, giving wrong information about the actual number of infected individuals. We calculate sensitivity indices and derive interesting epidemic-control information.

\textbf{Keywords}: Non-communicable disease, communicable disease, basic reproduction number, subcritical bifurcation, supercritical bifurcation, heterogeneous populations, risk-structured. 
\textbf{PACS}: 34, 
				37N25, 
				 92D30 

\end{abstract}

\section{Introduction}\label{sec:intro}
It is a fact that non-communicable diseases (NCDs) include cardiovascular disease, cancer, chronic respiratory disease, or diabetes, are the main cause of sickness and death worldwide \citep{WHO1}. Just in 2000 NCDs were responsible for 35 million deaths (about 60\% of all deaths around the world), and in 2020 these figures raised to 41 million death (i.e. about 71\% of deaths worldwide) \citep{WHO1}. NCDs are the result of a combination of non-reversible genetic and physiological factors but also of environmental and behavioral factors that may be reverted. Use of or exposure to tobacco, alcohol abuse, unhealthy diets, or physical inactivity \citep{GBD2015} are among these revertible factors, as well as air pollution and environmental contamination \citep{PruessUstuen2019}. NDCs are very common and, therefore, play a key role in the epidemiology of communicable or infectious diseases (CD) \citep{Coates2020}, \citep{Wong2021}.

The reproduction number, $R_0$, is a key quantity in the dynamics of a communicable disease. $R_0$ stands for the average number of secondary infections produced by an infected individual in a population made just of susceptible individuals \citep{Begon2002}. It is well known that $  R _0>1$ enables communicable diseases to become endemic. However  
$R _0<1$ does not always lead to the eradication of the communicable disease. This somewhat counter-intuitive fact is known as {\it subcritical bifurcation} \citep{Greenhalgh2000} (also often less properly named {\it backward bifurcation} \citep{Hadeler1997}). Mechanisms leading to subcritical bifurcations in epidemiological models are proposed in \citep{Gumel2012},
while general necessary and sufficient conditions for an epidemiological model to display a subcritical bifurcation
are obtained in \citep{Buonomo2015}. 
This phenomenon has important consequences from the viewpoint of epidemics control, since reducing $  R_0$ below $1$ may not be sufficient to avoid the disease endemic scenario. An instance of this unfortunate behavior is shown in the case of TB in India, \citep{Venturino2017}.

This work is aimed to analyze the interplay between NCDs and CDs. For this purpose, in Section \ref{sec:model} we set up a minimal model using one of the simplest transmission laws \citep{Begon2002} for CDs and the minimal number of epidemiological stages. The structure of the model presented herein can be seen as a simplified variant of risk-structured SIS models \citep{Keeling2007}, \citep{Gumel2012}, \citep{Kot2020}. In doing so, we isolate the {\it net effect} of the NCD/risk-structure on the behavior of the  CD. Thus, we disentangle the role of simple heterogeneity in the screened population from other processes  (see the discussion in Section \ref{sec:discussion}). 
In Section  \ref{sec:results} we analyze the model  and derive sufficient and necessary conditions enabling a  subcritical bifurcation. We discuss the results and its implications on the control of the CD in Section \ref{sec:discussion}.

\section{Model formulation}\label{sec:model}
We focus on the interplay between the CD and the NCD. Thus, we seek the minimal settings avoiding, for instance, demography processes. This approach yields a laboratory model that can be expanded to face more general settings. 

We assume that individuals affected by the NCD are somehow weaker to face the CD.  At time $t$ the population is partitioned into susceptible individuals $S(t)$, i.e.\ those that are affected neither by the CD nor the NCD, weakened individuals $W(t)$, i.e.\ those that suffer from the NCD but are not infected by the transmissible disease, and individuals infected by the CD, $I(t)$,  regardless of whether they are weakened or not. 

\begin{figure}[h!]
	\centering
		\includegraphics[width = 9cm]{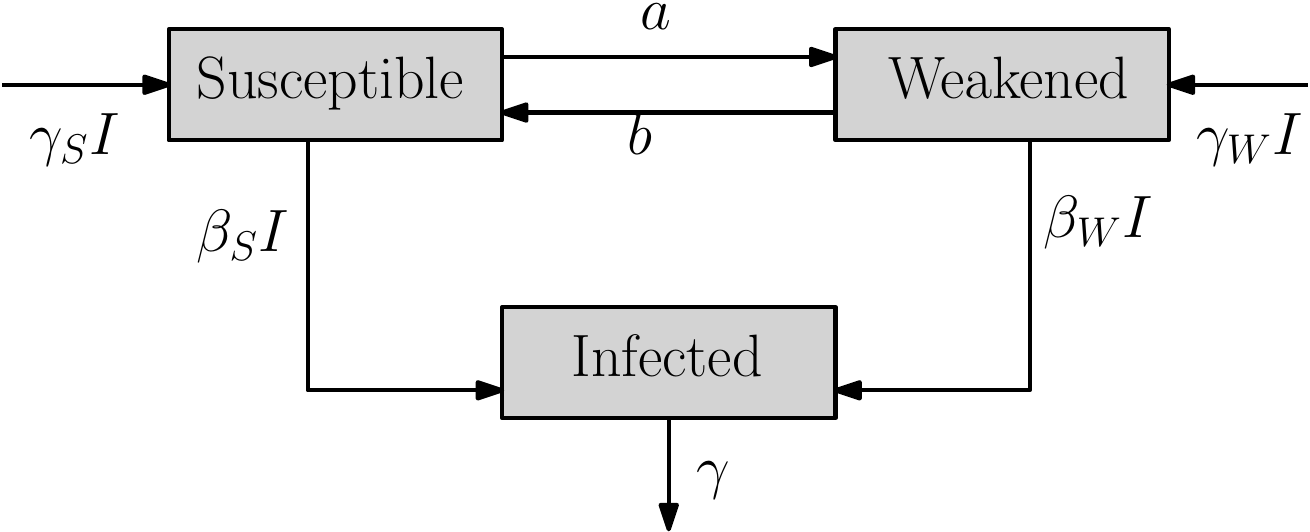}
	\caption{Flow diagram associated to system (\ref{eq:model}).}
	\label{fig:flow}
\end{figure}

The model is built with ordinary differential equations. Next, we define the rates at which individuals move from one compartment to each other, see Figure \ref{fig:flow}.  
We assume that susceptible individuals become weakened at a constant rate $a$ and get out of the weakened class also at a constant rate $b$. It may refer to many different processes. For instance, for diseases not directly transmissible through contact (for instance, smoking) still the ``observation'' of
the other people's opposed behaviors may influence
both susceptible to take the same habit, or conversely to suggest ``addicted'' people (i.e.\ $W$ individuals) to leave stop smoking. On the other hand, external dynamics may impoverish/take out of poverty people, which is also a risk factor in front of a CD. Note that the transmission of the NCD could still be formulated considering more sophisticated functional terms \citep{Raimundo2018}, \citep{Opoku2021}, but we disregard this approach here to keep the model formulation minimal as mentioned. The dynamics of the transmissible disease is somehow similar to (but not exactly)  the classical SIS model \citep{Begon2002} with density-dependent transmission.  Susceptible and weakened individuals are assumed to behave differently concerning the CD so that the transmission rates $\beta_S$ and $\beta_W$ differ from each other. We assume that infected individuals recover and become immune at rate $\gamma$. Note that we do not know whether an infected individual suffers or not from the NCD. Thus, we do not care about recovered individuals (that leave the model). We further consider that susceptible (in front of the CD)  individuals are introduced at rates $\gamma_S I$ and $\gamma_W I$ in the corresponding compartment $S$ and $W$, so that the population size is kept constant.

The analysis can be done {\it mutatis mutandi} also by considering frequency-dependent transmission. 

The ordinary differential equations system produced by the above-stated hypotheses consists of two coupled submodels: one describing the communicable disease and another one that describes the non-communicable disease.
The combined model reads as follows
\begin{equation}\label{eq:model}
\left\{
\begin{array}{l}
	S' = -aS+bW-\beta_SSI+\gamma_SI,\\
	\\
	W' =  aS - bW - \beta_WWI + \gamma_WI,\\
	\\
	I' = \beta_SSI + \beta_WWI - \gamma I, \\
\end{array}\right.
\end{equation}
where $\gamma_S+\gamma_W = \gamma$.  Note that the total population size $N(t):=S(t)+W(t)+I(t)=N$ remains constant, as $N'(t)=0$. \newline

For sake of completeness, let us revisit these two well known models: the SIS and the model for non-communicable diseases.

\paragraph{SIS model (CD).} If susceptible individuals are all of the class, system (\ref{eq:model}) simplifies into 
\begin{equation}\label{eq:SIS}
\left\{
\begin{array}{l}
	S' = -\beta_SSI+\gamma_SI,\\
	\\
	I' = \beta_SSI - \gamma_SI,\\
\end{array}\right.
\end{equation}
where $S(t)+I(t)=N$ remains constant over time. It is nothing but the classical SIS model with density-dependent transmission \citep{Begon2002}. A straightforward analysis reveals that system (\ref{eq:SIS}) possesses two equilibrium points: the trivial equilibrium $(S^*,I^*) = (N,0)$ (no infected individuals) and a the endemic-disease equilibrium 
\begin{equation}\label{eq:com-no-triv}
	(S^*,I^*)=\left(\dfrac{\gamma_S}{\beta_S}, \, N-\dfrac{\gamma_S}{\beta_S}\right)
\end{equation}
The well known {\it basic reproduction number} 
\begin{equation}\label{eq:R0sis}
	 {R}_{0,S}=\dfrac{\beta_S}{\gamma_S}N
\end{equation}
determines whether the disease-free equilibrium ($ {R}_{0,S}<1$) or the endemic-disease equilibrium ($ {R}_{0,S}>1$) is the global attractor of system (\ref{eq:SIS}).

\paragraph{Non-communicable disease.} In the absence of the infectious disease ($I(t)=0$), system (\ref{eq:model}) reduces to system 
\begin{equation}\label{eq:NCD}
\left\{
\begin{array}{l}
	S' = - aS + bW,\\
	\\
	W' = aS - bW,\\
\end{array}\right.
\end{equation}
where $S(t)+W(t)=N$ remains constant over time. We will discuss in section \ref{sec:discussion} further extensions
of this submodel. 
We assume $N>0$ (there are individuals) so that there exists a non trivial equilibrium point 
\begin{equation}\label{eq:non-com-no-triv}
	(S^*,W^*)=\left(\dfrac{b}{a+b}N, \, \dfrac{a}{a+b}N\right)
\end{equation}
where $b/(a+b)$ and $a/(a+b)$ are the fraction of susceptible and weakened individuals within the entire population $N$. It is straightforward that the nontrivial equilibrium (\ref{eq:non-com-no-triv}) is a global attractor. This feature can be interpreted as the non-communicable disease being structural to the population.

\section{Results}\label{sec:results} 
In this section, we analyze the long term behavior of the solutions of system (\ref{eq:model}), that is, the so-called equilibrium points and their stability.

A first step consists of showing that the model is well behaved, that is,
\begin{theorem} The solutions of system (\ref{eq:model}) are bounded from above and the non negative cone is forward invariant. 
\end{theorem} 
\noindent\textbf{Proof}: All the solutions of system (\ref{eq:model}) are bounded since the total population size is kept constant because $S'(t)+E'(t)+I'(t)=0$. 
The invariance of the non negative cone 
$$\bar{\mathbb{R}}_+^3:=\left\{(S,W,I)\in \mathbb{R}^3;\, S\geq0,\,E\geq0,\,I\geq 0\right\}$$ 
is equivalent to prove that any solution with initial values on the boundary can not become negative as the time 
flows. For instance, assume that $W(t_0)=0$ and $S(t_0)\cdot I(t_0)\neq 0$. It follows that $W(t_0)' = aS(t_0)+\gamma_W I(t_0)>0$, so that the corresponding solution grows towards positive values, and the same holds assuming $S(t_0)=0$,  $W(t_0)\cdot I(t_0)\neq 0$. 

Assume now that $I(t_0)=0$ and  $S(t_0)\cdot W(t_0)\neq 0$. Then $I'(t_0)=0$ regardless of the value of $S(t_0)$ and $W(t_0)$. There are no infected individuals and there will be none. The solution of the system evolves constrained by $S(t_0)+W(t_0) = N=cte$, that is to say that system (\ref{eq:model})  is reduced to system (\ref{eq:NCD}) and its solution converge to (\ref{eq:non-com-no-triv}).
$\blacksquare$\\

As we have already said, the non-communicable disease is supposed to be inherent to that population. The first result consists of determining  conditions so that an outbreak of the communicable disease lead to an endemic disease scenario. In other words, we seek conditions enabling the semitrivial equilibrium point 
\begin{equation}\label{eq:W0}
E_0^* = 
(S_0^*,W_0^*, I_0^*)=\left(\dfrac{b}{a+b}N, \, \dfrac{a}{a+b}N, \, 0\right)
\end{equation}
to be asymptotically stable (communicable-disease-free state) or unstable (endemic communicable disease scenario).
Note that trivial equilibrium point $E_0^*$ consists of the components of the nontrivial equilibrium
(\ref{eq:non-com-no-triv}) of system (\ref{eq:NCD}) along with $0$ infected individuals in the third entry.

\begin{theorem} \label{prop:semitrival} Consider system (\ref{eq:model}), the disease free equilibrium $E_0^*$ given by (\ref{eq:W0}) and the reproductive number $ R_0$ defined by 
\begin{equation}\label{eq:r0}
  R_0 = \dfrac{b\beta_S + a\beta_W}{(a+b)(\gamma_S+\gamma_W)}N.
\end{equation}
Then $E_0$  is unstable if $  R_0<1$ and locally asymptotically stable if $  R_0>1$.
\end{theorem} 
\noindent\textbf{Proof}: It follows from a standard analysis of the sign of the eigenvalues of the Jacobian matrix of the flow of system (\ref{eq:model}). $\blacksquare$\\

Thus the CD free equilibrium (\ref{eq:W0}) is unstable 
if $ R_0>1$, which leads to a CD endemic scenario.
On the other hand, $R_0<1$  implies that the CD free equilibrium (\ref{eq:W0}) is locally asymptotically stable, meaning that  any CD outbreak will fade out (at least if the number of infected individuals is small enough). These results provide us with valuable but incomplete information. Namely:
\begin{enumerate}
	\item What are the conditions leading the CD free scenario to be globally asymptotically stable? I.e., what are the conditions ensuring that CD free scenario will be achieved regardless of the strength of a potential outbreak?
	
	\item What role does the NCD play in the dynamics of communicable disease? 
\end{enumerate}
In other words, we are interested in the structure of the set of the positive equilibrium points of system (\ref{eq:model}) and its stability. Positive equilibrium points are the component-wise positive solutions to system 
\begin{equation}\label{eq:equil}
\left\{
\begin{array}{l}
	0 = -aS+bW-\beta_SSI+\gamma_SI,\\
	\\
	0= aS - bW - \beta_WWI + \gamma_WI,\\
	\\
	0 = \beta_SSI + \beta_WWI - \gamma I,\\
\end{array}\right.
\end{equation}
where $\gamma = \gamma_S+\gamma_W$. Let us assume that $I(t_0)\neq 0$, since otherwise system (\ref{eq:equil}) reduces to system (\ref{eq:NCD}). Weakened individuals behave differently from susceptible individuals in front of the CD, being plausible $\beta_W\geq\beta_S$ and $\gamma_W\leq\gamma_S$ with at least one of the inequalities being strict. We assume $\beta_W>\beta_S$ through the manuscript, since $\gamma_S$ and $\gamma_W$ are not proper recovery rates.

\begin{figure}[h!]
	\centering
		\includegraphics[width = \linewidth]{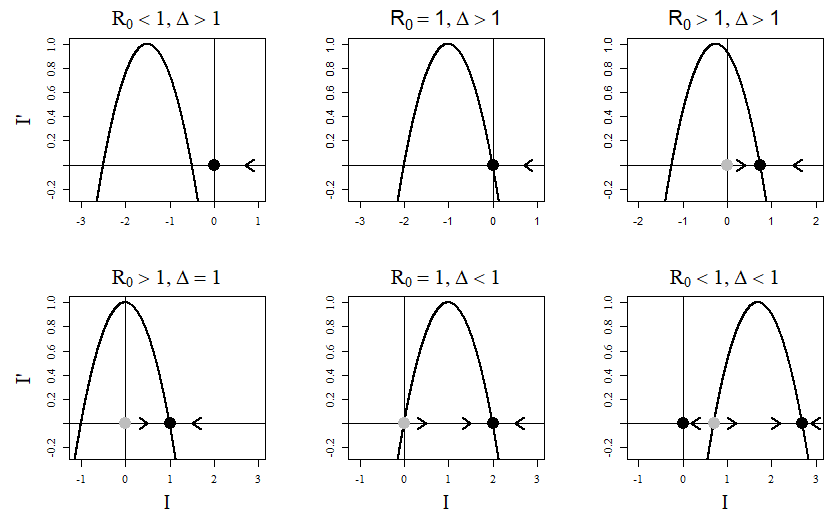}
	\caption{
Each panel displays the parabola defined by  equation (\ref{eq:I}) in the $I$-$I'$ plane for different combinations of $R_0$ and $\Delta$. Solid dots are the feasible equilibrium points that are asymptotically stable (black) and unstable (gray).}
	\label{fig:parabolas}
\end{figure}

Using the fact that the total population $S(t)+I(t)+W(t) = N$ is constant, direct calculations yield that the number of infected individuals is the solution of a quadratic polynomial equation
\begin{equation}\label{eq:I}
\Psi(I) = \alpha_2 I ^2 + \alpha_1 I+\alpha_0 = 0
\end{equation}

where
\begin{equation}\label{eq:coefs}
\begin{array}{c}
\alpha_2 = -\dfrac{\beta_S\beta_W}{\beta_W-\beta_S}<0, \\ \\
\alpha_1 = -\dfrac{\beta_S(b+\gamma_W)+\beta_W(a+\gamma_S)-\beta_S\beta_W N}{\beta_W-\beta_S},\\
\\
\alpha_0 = -\dfrac{(\gamma_S+\gamma_W)(a+b)-(b\beta_S+a\beta_W)N}{\beta_W-\beta_S}.
\end{array}
\end{equation}
Note that $\beta_W>\beta_S$, since weakened individuals are weaker in front of the CD. It is immediate calculate the number of infected individuals $I^*$ (if any) at equilibrium. Then, $S^*$ and $W^*$, the number of susceptible and weakened individuals at equilibrium can be calculated from $I^*$ according to 
\begin{equation}\label{eq:is}
\begin{array}{c}
I_{\pm}^* = \dfrac{-\alpha_1\pm \sqrt{\alpha_1^2-4\alpha_0\alpha_2}}{2\alpha_2}, \\
S_{\pm}^* = \dfrac {\beta_W}{\beta_W-\beta_S}\left( N-\dfrac {\gamma_S + \gamma_W}{\beta_W} - I_{\pm}^*\right), \\
\\
W_{\pm}^* =  \dfrac {\gamma_S + \gamma_W - \beta_S S_{\pm}^*} {\beta_W}.
\end{array}
\end{equation}
Note that condition $\alpha_1^2-4\alpha_0\alpha_2<0$ yields no infected individuals, so that we assume from now on that $\alpha_1^2-4\alpha_0\alpha_2\geq 0$.
Direct calculations yield
\[
\alpha_0=0 \Leftrightarrow   R_0=1, \quad
\alpha_0<0 \Leftrightarrow   R_0<1, \quad
\alpha_0>0 \Leftrightarrow   R_0>1
\]
along with
\[
\alpha_1=0 \Leftrightarrow \Delta = 1, \quad
\alpha_1>0 \Leftrightarrow \Delta>1, \quad
\alpha_1<0 \Leftrightarrow \Delta>1,
\]
where 
	\begin{equation}\label{eq:delta}
	\Delta = 
	\left(\dfrac{a+\gamma_S}{\beta_S} + \dfrac{b+\gamma_W}{\beta_W}\right)\dfrac{1}{N} 
	\end{equation}
It is not difficult to classify all the possible qualitatively different outcomes in terms of sign of $\alpha_0$ and $\alpha_1$ (given that $\alpha_2<0$). That is to say, in terms of the value of $ R_0$ and $\Delta$ relative to $1$.  Figure \ref{fig:parabolas} sketches the interesting cases. 

Panels in Figure \ref{fig:parabolas} display $I'=\Psi(I)$; that is to say, the intercept with the horizontal axis is the amount of infected individuals at equilibrium (since $I'=0$). Note that  the sign of $I'(t)$ determines whether the number of infected individuals $I(t)$ increases ($I'>0$) or decreases ($I'<0$), which yields the stability of the equilibrium points. For instance, consider the right bottom panel: let $I(t,I_0)$ be the solution to  $I'=\Psi(I)$ such that $I_0=I(t_0,I_0)$, for $I_0$ larger than the most right equilibrium point (say $I^{*}$). It is apparent $I'(t_0,I_0)<0$, so that $I(t,I_0)$ is strictly decreasing for any $t\in(t_0,t_1)$. Also, $I^*<I(t,I_0)$ for any  $t\in(t_0,t_1)$. Let $t_1$ define the maximal interval where $I(t,I_0)$ is decreasing. $I(t,I_0)$ is bounded from below, strictly decreasing and continuous. Then, there exists 
\begin{equation}\label{eq:limit}
    \hat{I}=\lim_{t\to\infty}I(t,I_0).
\end{equation}
The uniqueness of solutions precludes $\hat{I}<I^*$. Also, $\hat{I}>I^*$ implies that $I'(t,I_0)<\xi<0$ for all $t\in(t_0,t_1)$, that is a contradiction with the existence of limit (\ref{eq:limit}). Then, $\hat I = I^*$, which concludes the proof. 

The stability of all the other equilibrium points in Figure \ref{fig:parabolas} follows reasoning as before.\\\newline

\begin{figure}[h!]
	\centering
		\includegraphics[width = 12cm]{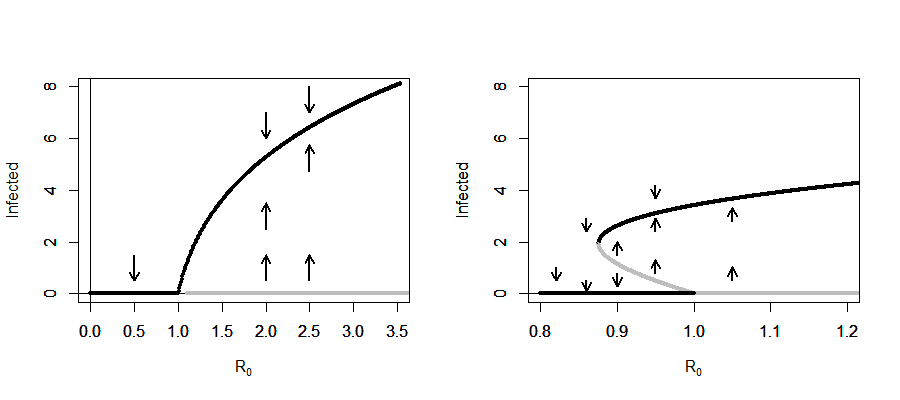}
	\caption{
Bifurcation diagrams plotting the total number of infected individuals at equilibrium for different values of $  R_0$. The left panel displays a supercritical bifurcation and 
	the right panel displays a subcritical bifurcation. 
	In gray unstable equilibrium points and in black asymptotically stable equilibrium points. 
	Parameter values: , $\beta_W=12.1$; $\beta_S=2.6$; $\gamma_S=8.4$; $\gamma_W=33.3$, $a\in[0.001, 10]$
}
	\label{fig:bifurcation}
\end{figure}

Essentially, three scenarios are possible: the global CD free scenario (left and central panels in the first row of Figure \ref{fig:parabolas}), the global endemic CD scenario (right panel in the first row and left and central panels at the second row of Figure \ref{fig:parabolas}), and a third intermediate one that predicts either endemic disease or disease free scenarios depending on the initial amount of infected individuals (right panel of the second row of Figure \ref{fig:parabolas}). These features are better shown with a bifurcation diagram, being $R_0$ the bifurcation parameter, see Figure \ref{fig:bifurcation}. In this context, $\Delta$ is the so-called direction of bifurcation, so that $\Delta<1$ leads to a subcritical (or backward) bifurcation (right panel in Figure \ref{fig:bifurcation}) and $\Delta>1$ yields a 
supercritical (or forward) bifurcation (left panel in Figure \ref{fig:bifurcation}).

For $R_0<1$ the solution $I^*=0$ is locally asymptotically stable, i.e.\ the presence of a small number of infected individuals is not enough to trigger an epidemic disease state, and the infected population will fade away regardless of the value of $\Delta$. However, due to management decisions or natural causes, the values of the parameters involved in the expression of $R_0$ may change and increase $R_0$ so that it crosses the threshold value $R_0=1$. In such a case, the solution $I^*=0$ is destabilized, which means that the communicable disease becomes endemic even if there is a little initial amount of infected individuals. 

The quantity $\Delta$ plays a key role when $I^*=0$ is stable ($R_0<1$): namely if $\Delta>1$ then  $I^*=0$ is globally asymptotically stable, which means that any epidemic outbreak will fade away regardless of the initial number of infected individuals. On the contrary, $\Delta<1$ implies that there exists $R_0^*<1$ such that for each  $R_0^*<R_0<1$ there exist a threshold number of infected individuals given by $I_-^*$ (see equation (\ref{eq:is})) such that beyond it the disease becomes endemic and stabilize at $I_+^*$ (see equation (\ref{eq:is})). For $\Delta<1$ and $R_0<R_0^*$ $I^*=0$ becomes globally asymptotically stable. 

The stability of all the other equilibrium points in Figure \ref{fig:parabolas} follows reasoning as before.\\\newline

\section{Discussion}\label{sec:discussion}

We next discuss our finding on system (\ref{eq:model}). Its apparent simplicity allows, instead, to emerge relevant features. First of all, we account for the possibility of subcritical bifurcation  under minimal settings, see Section  \ref{sec:subcritical}. We also analyze the consequences of not considering explicitly the NCD  in Section  \ref{sec:no-NCD}. Then, in Section  \ref{sec:sensitivity}, we focus on the effect of control strategies (via modifying the coefficients of the system). In particular, we calculate the sensitivity indices, we derive bounds for these indices, we focus on the effect of modifying several coefficients at once, and we reveal possible unexpected consequences when trying to handle disease outbreaks. 

\subsection{Subcritical bifurcation} \label{sec:subcritical}
Regardless of the approach (CD vs NCD or CD risk structure) system (\ref{eq:model}) may undergo a subcritical bifurcation, which is against the $R_0$-dogma \citep{Reluga2008} that states that $R_0>1$ leads to disease endemicity while $R_0<1$ induces disease eradication. Usual causes of subcritical bifurcation are the use of imperfect vaccine \citep{Brauer2004}, \citep{Gumel2021}, structured immunity \citep{Reluga2008} or  exogenous re-infection in TB disease \citep{Feng2000}. 

In \citep{Gumel2012} several other biological or epidemiological mechanisms are proposed 
such as vaccine-induced immunity waning at a slower rate than natural immunity, disease-induced mortality in vector-borne diseases, and differential susceptibility in risk-structured models (related to the latter, see  \citep{Keeling2007} and \citep{Kot2020}). 

Subcritical bifurcations can be found also in co-infection by an opportunistic disease model \citep{Marva2018}, where two communicable diseases were considered  (one of them with saturating treatment rate \citep{Martcheva2015}).

A salient feature of the model proposed and analyzed here lies in its ability to undergo a subcritical bifurcation while not incorporating any of the above-mentioned mechanisms. Thus, we show that subcritical bifurcations in epidemiology are not such a rare occurrence. On the contrary, plain heterogeneity in the CD susceptible class is enough to make a subcritical bifurcation possible. 



Thus, system (\ref{eq:model}) shows that subcritical bifurcations may occur in epidemic models simply by considering the dynamics associated with a heterogeneous population, which can be seen as a common factor in the above-mentioned models. 



\subsection{What if the non-communicable disease is not explicitly  considered?} \label{sec:no-NCD}
Thus, let us assume that there is no weakened individuals compartment so that the CD follows the simplest model  (\ref{eq:SIS}). Even if we do not consider an explicit compartment, the NCD is present in the population, and the coefficients of system (\ref{eq:SIS}) must reflect in some way this fact. It is reasonable assuming that the NCD-induced population heterogeneity will be captured by any reasonable sampling procedure performed to estimate the coefficients of the model by weighting the corresponding transmission ($\beta_S$ and $\beta_W$) and recovery ($\gamma_S$ and $\gamma_W$) coefficients. This hypothesis is equivalent to assuming that the dynamics associated with the NCD has already achieved an equilibrium, which is the usual assumption when dealing with time-scale systems \citep{Auger2008} (also known as {\it quasi-steady-state approximation} \citep{Garde2020}) that has been used in co-infection by an opportunistic disease models \citep{Marva2015}, \citep{Marva2018}, where both diseases are transmissible. Thus a fraction $b/(a+b)$ of the total population is free of the NCD and the remaining fraction  $a/(a+b)$ is not. Then, to obtain a fair comparison,
transmission and the recovery rates are set to 
\begin{equation}\label{eq:averaged}
\beta_S\dfrac{b}{a+b}+\beta_W\dfrac{a}{a+b}, \qquad \gamma_S\dfrac{b}{a+b}+\gamma_W\dfrac{a}{a+b},
\end{equation}
respectively, which yield the corresponding basic reproductive number:
\begin{equation}\label{eq:R0noNCD}
\widehat{  R}_0 = \frac{b\beta_S+a\beta_W}{b\gamma_S+a\gamma_W}N, \qquad \qquad 
(\widehat S^*,\,\widehat I^*)=N\left(\frac{1}{\widehat{  R}_0},\, N- \frac{1}{\widehat{  R}_0}
\right)
\end{equation}
Direct calculations yield
\begin{equation}\label{eq:R0ratio}
\frac{\widehat{  R}_0}{  R_0} = 1+\frac{a\gamma_S+b\gamma_W}{b\gamma_S+a\gamma_W}.
\end{equation}
That is, the ratio (\ref{eq:R0ratio}) is always larger than $1$, implying that explicit consideration
of the NCD dynamics in the model does matter. Neglecting its effect leads to overestimating the basic reproductive number and, thus, i) thinking of an endemic disease scenario that may be not real and ii) overestimating the number of infected individuals at equilibrium. 

\begin{figure}[h!]
	\centering
		\includegraphics[width = \textwidth]{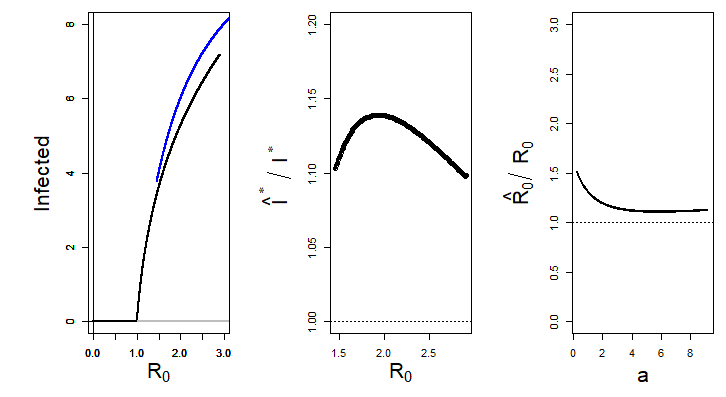}		
	\caption{Left panel: bifurcation diagram of the number of infected individuals at equilibrium of the classical SIS model (\ref{eq:SIS}, in blue, top curve) 
with transmission rate $\beta = b\beta_S/(a+b)+a\beta_W/(a+b)$ and recovery rate  $\gamma = b\gamma_S/(a+b)+a\gamma_W/(a+b)$ 
and  system (\ref{eq:model}) in black, bottom curve. The bifurcation parameter is $\hat R_0$ and $R_0$, respectively, for $N=12, b=35.4, \beta_S = 1, \beta_W=9.1, 
\gamma_S=8.4, \gamma_W=15.3$ and $a\in[0.1,10]$. Note that expression (\eqref{eq:R0ratio}) and the parameters values explain the aparent gap in $\widehat  I^*$. 
Central panel, the ratio $\widehat  I^*/I^*$. 
Right panel: the ratio $\widehat R_0/R_0$.
}
\label{fig:SIS-vs}
\end{figure}

 
Figure \ref{fig:SIS-vs} displays the bifurcation diagram of 
the total amount of infected individuals at equilibrium 
$I^*$ (in black, bottom line) and $\widehat I^*$ (in blue, upper line) versus $ { \widehat R}_0$ and $  R_0$,
that both appear in the horizontal axis. 
 

\subsection{Sensitivity analysis, sensitivity indices and epidemic control}\label{sec:sensitivity}
Finally, we accomplish a sensitivity analysis of the outcome of the model to the parameters of the model. 

In Section \ref{sec:results} we have shown that the long term behavior of the model can fully be described in terms of $R_0$ and $\Delta$. We first examine the expressions of $R_0$ and $\Delta$. Next, we calculate the corresponding sensitivity indices \citep{Chitnis2008}. Then we have drawn concussion useful for  control purposes.\\\newline

Grouping terms in the expression of $\Delta$ (see equation (\ref{eq:delta})) yields 
	\begin{equation}\label{eq:delta2}
	\begin{array}{rl}
	\Delta & = \left(\dfrac{\gamma_S}{\beta_S} + \dfrac{\gamma_W}{\beta_W}\right)\dfrac{1}{N} + 
	\left(\dfrac{a}{\beta_S} + \dfrac{b}{\beta_W}\right)\dfrac{1}{N} \\
	& \\
	& = 	\underbrace{\dfrac{1}{  R_{0,S}} +  \dfrac{1}{  R_{0,W}}}_{\text{Block 1}} 
	+ 
	\underbrace{\left(\dfrac{a}{\beta_S} + \dfrac{b}{\beta_W}\right)\dfrac{1}{N}}_{\text{Block 2}}
	\end{array}
  \end{equation}
	
\begin{figure}[h!]
	\centering
		\includegraphics[width = 12cm]{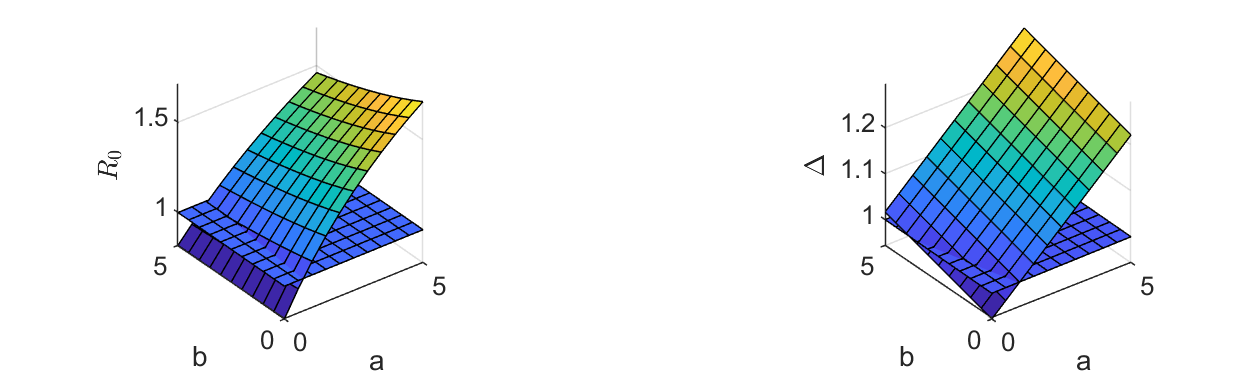}\\
		\includegraphics[width = 12cm]{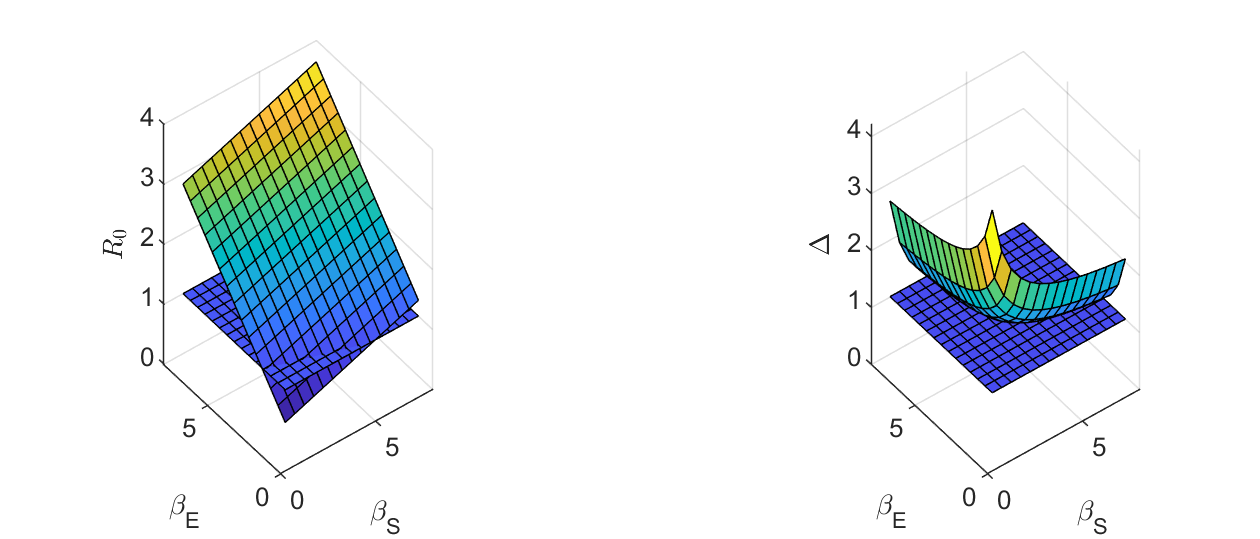}
	\caption{Left (right, respt.) column, $R_0$ ($\Delta$, respt.) as function of coefficients $a,\,b$ (top) and $\beta_W,\,\beta_S$ (bottom). All the panels display also the threshold plane $R_0=1$ (left column) and $\Delta = 1$ (right column). } 
	\label{fig:sensitivity}
\end{figure}

Interestingly, the first block includes the basic reproduction numbers corresponding to either no weakened class (all individuals are susceptible) or no susceptible class (all individuals are weakened, i.e., the analogous case with other values for $\gamma$ and $\beta$). In contrast, the second block includes the ratio of the rates at which individuals leave the susceptible class ($a/\beta_S$) or the weakened class ($b/\beta_W$). Two conclusions can be drawn from (\ref{eq:delta2}): on the one hand, $\Delta$ depends linearly on $a$ and $b$ (see the top right panel of Figure \ref{fig:sensitivity}). Therefore, $\Delta$ changes linearly with these parameters. On the other hand, $\Delta$ depends non-linearly on the corresponding basic reproduction numbers or, ultimately, on the transmission rates (see the bottom right panel of Figure \ref{fig:sensitivity}). 

Concerning $  R_0$, it depends linearly on $\beta_S$ and $\beta_W$ (see expression (\ref{eq:r0}) and the bottom left panel of Figure \ref{fig:sensitivity}). On the contrary, $  R_0$ depends non-linearly on $a$ and $b$, although the effect is quasi-linear (see expression (\ref{eq:r0}) and the top left panel of Figure \ref{fig:sensitivity}).\\\newline

\subsubsection{Sensitivity indices}\label{sec:sensindex}
In avoiding the communicable disease becoming endemic (or promoting endemicity) we must control the values of $R_0$ and $\Delta$ (see Figure 3). It is useful to know the relative importance of the parameters involved in the expressions of $R_0$ and $\Delta$, so we can choose which of them must be changed when developing intervention strategies.

The normalized forward sensitivity index of a variable, $u$, that depends differentiably on a parameter $p$ is defined in \citep{Chitnis2008} as:
\begin{equation}\label{eq:sensit}
    \gamma_p^u:=\frac{\partial u}{\partial p}\times\frac{p}{u} = \frac{\frac{\partial u}{\partial p}}{\frac{u}{p}} 
\end{equation}
For the convenience of the reader we sketch the derivation of the sensitivity index (\ref{eq:sensit}) in Appendix \ref{sec:app}. Its interpretation is as follows. 
When increasing (or decreasing) $p_0$ by $\epsilon$ (meaning increasing  $p_0$ to $p_0+\epsilon p_0=(1+\epsilon)p_0$)  $u$  increases (or decreases) by $\epsilon\left.\gamma_p^u\right|_{p = p_0}$ times $u(p_0)$. Let us underline that this interpretation is {\it local} and {\it approximated} in the same sense the Taylor's expansion is so (see Appendix (\ref{sec:app})).

We have used (\ref{eq:sensit}) to derive the analytical expression for the sensitivity index of $R_0$ and $\Delta$, defined by (8) and (13) respectively, to each of the six parameters considered in our model. We show the corresponding expressions in Table \ref{tab:indices}. 
   
\begin{table}[!ht]
    \centering
    \begin{tabular}{|c|c|c|}
    \hline
          & $R_0$ & $\Delta$ \\ \hline \hline
          && \\ 
        $a$ & $\frac{ab}{a+b}\frac{\beta_w-\beta_s}{b\beta_s+a\beta_w}$ & $\frac{a\beta_w}{(a+\gamma_s)\beta_w+(b+\gamma_w)\beta_s}$  \\ 
         && \\ \hline 
         && \\
        $b$ & $\frac{ab}{a+b}\frac{\beta_s-\beta_w}{b\beta_s+a\beta_w}$ & $\frac{b\beta_s}{(a+\gamma_s)\beta_w+(b+\gamma_w)\beta_s}$\\ 
         && \\ \hline 
         && \\
        $\beta_s$ & $\frac{b\beta_s}{b\beta_s+a\beta_w}$ & $-\frac{(a+\gamma_s)\beta_w}{(a+\gamma_s)\beta_w+(b+\gamma_w)\beta_s}$ \\ && \\\hline 
         && \\ 
        $\beta_w$ &$\frac{a\beta_w}{b\beta_s+a\beta_w}$ & $-\frac{(b+\gamma_w)\beta_s}{(a+\gamma_s)\beta_w+(b+\gamma_w)\beta_s}$ \\ && \\\hline 
         && \\ 
        $\gamma_s$ & $-\frac{\gamma_s}{\gamma_s+\gamma_w}$ & $\frac{\gamma_s\beta_w}{(a+\gamma_s)\beta_w+(b+\gamma_w)\beta_s}$\\ 
        && \\\hline 
         && \\
        $\gamma_w$ & $-\frac{\gamma_w}{\gamma_s+\gamma_w}$ & $\frac{\gamma_w\beta_s}{(a+\gamma_s)\beta_w+(b+\gamma_w)\beta_s}$\\ &&\\\hline
    \end{tabular}
    \caption{Sensitivity indices of $R_0$ (8) and $\Delta$ (13) to recovery and transmission rates of the NCD and the transmissible disease considered in our model.}
    \label{tab:indices}
\end{table}

Assume now that $u$ depends on parameters $p_1,\cdots,p_n$. Without lost of generality, we assume that all the  parameters vary simultaneously by $\epsilon$. A direct application of the generalized Taylor's expansion yields that the corresponding sensitivity index is 
\begin{equation}
    \label{eq:variation several parameters}
    \sum_{i=1}^{n}\gamma_{p_i}^u
\end{equation}
That is to say that the sensitivity indices are additive, but (or and) the sign of each sensitivity index matters. 
Note that when the sum of the different indices is 1 (or -1) the variation caused in the variable $u$ is the same variation introduced in the parameters times the corresponding sensitivity index.

Most of the expressions of these sensitivity indices are complex so it is not possible to set an order from most sensitive to least sensitive without evaluating them at some baseline parameter values. However, some general conclusions can be drawn.

\subsubsection{Implications for managing disease outbreaks}\label{sec:implications}
Epidemiologists look at $R_0$ at the beginning of epidemic outbreaks \citep{Boonpatcharanon2021}, \citep{Keeling2007} (but see also \citep{Delamater2019} for a less theoretical approach). $R_0$ depends on the parameters of the model and a key question is that of ascertaining which coefficients modify to get the larger change in $R_0$ with minimum effort.

We next derive information from the expression of the sensitivity indices useful for disease managers (see Table \ref{tab:indices}) that may be useful for disease managers. All the mathematical relations follow straightforward from the expressions gathered in Table \ref{tab:indices}. We assume that $a,\,b,\,\beta_s,\,\beta_w,\,\gamma_s,\,\gamma_w > 0$.

\textbf{Bounds for the sensitivity indices} On the one hand, most of the expressions for the sensitivity indices are fractions in which the numerator is one of the summands of the denominator. As all the parameters are positive quantities, the absolute value of these indices is less than one. More specifically, Table \ref{tab:maximum values} shows infimum and supremum values of sensitivity indices of $R_0$ and $\Delta$.
\begin{table}[]
    \centering
    \begin{tabular}{|c|c|c|c|c|}
    \hline
        & \multicolumn{2}{|c|}{Sensitivity indices of $R_0$}& \multicolumn{2}{|c|}{Sensitivity indices of $\Delta$}\\\hline
       &Infimum value&Supremum value&Infimum value&Supremum Value\\\hline \hline
       $a$  & - & - & 0 & 1 \\\hline
         $b$& - & - & 0 & 1\\\hline
        $\beta_s$ & 0 & 1 & -1 & 0 \\\hline
         $\beta_w$ & 0 & 1 & -1 & 0\\\hline
         $\gamma_s$ & -1 & 0 & 0 & 1 \\\hline
        $\gamma_s$ & -1 & 0 & 0 & 1\\\hline
    \end{tabular}
    \caption{Infimum and supremum values of sensitivity index of $R_0$ and $\Delta$. The cells are empty when no meaningful bounds can be provided.}
    \label{tab:maximum values}
\end{table}

As shown in Table \ref{tab:maximum values}, the supremum value of the sensitivity indices is $1$. So, it is not possible to change $R_0$ (or $\Delta$) by an amount bigger than the change  $\epsilon$ introduced in the parameter times $R_0$ (or $\Delta$).The interpretation of the infimum value $-1$ is the same, but in that case, the modification introduced in the parameters and the change in $R_0$ (or $\Delta$) have the opposite direction.
In addition, note that we are considering a Taylor expansion to first order, so no big changes can be targeted.

\textbf{Comparing communicable disease management strategies}
Tables \ref{tab:indices} and \ref{tab:maximum values} allow us to compute how much  modifying a {\it single} coefficient of the model makes $R_0$ or $\Delta$ vary. However, it is possible to act on {\it more than one} coefficient of the model at once. Table \ref{tab:sums} gathers combinations of parameters to be modified simultaneously and equally that produce minimal, none, or maximal responses on $R_0$ and $\Delta$.

\begin{table}[h!]
    \centering
    \begin{tabular}{|c|c|}
    \hline
        $R_0$&$\Delta$ \\\hline \hline 
       &\\$ \gamma_{\beta_s}^{R_0}+\gamma_{\beta_w}^{R_0}=1$  & $\gamma_{\beta_s}^{\Delta}+\gamma_{\beta_w}^{\Delta}=-1$  \\&\\\hline
       &\\$\gamma_{\gamma_s}^{R_0}+\gamma_{\gamma_w}^{R_0}=-1$ & -- \\&\\\hline
   &\\ $\gamma_{\gamma_s}^{\Delta}+\gamma_{\gamma_w}^{\Delta}
    +\gamma_{a}^{\Delta}+\gamma_{b}^{\Delta}=-1$ & $\gamma_{\gamma_s}^{\Delta}+\gamma_{\gamma_w}^{\Delta}
    +\gamma_{a}^{\Delta}+\gamma_{b}^{\Delta}=1$ \\&\\\hline
       &\\ $\gamma_{a}^{R_0}+\gamma_{b}^{R_0}=0$ & -- \\&\\\hline
       &\\  $\gamma_{\beta_s}^{R_0}+\gamma_{\beta_w}^{R_0}+
    \gamma_{\gamma_s}^{R_0}+\gamma_{\gamma_w}^{R_0}+\gamma_{a}^{R_0}+\gamma_{b}^{R_0}=0$ & $\gamma_{\beta_s}^{\Delta}+\gamma_{\beta_w}^{\Delta}+
    \gamma_{\gamma_s}^{\Delta}+\gamma_{\gamma_w}^{\Delta}+\gamma_{a}^{\Delta}+\gamma_{b}^{\Delta}=0$ \\&\\\hline
    \end{tabular}
    \caption{Expressions of sums of indices when different combinations of parameters are simultaneously modified producing minimal, none or maximal response on $R_0$ and $\Delta$.}
    \label{tab:sums}
\end{table}

We next examine how the expressions gathered in Table \ref{tab:sums} can be used to decide on epidemics management strategies.\\

A key question from the point of view of managing a CD that of deciding to act either on a target population (for instance, weakened individuals) or equally on all the susceptible individuals regardless of their status. More specifically, the question is: What will produce a larger change in $R_0$, an effect $\epsilon_1$ applied only on (say) $\beta_w$, or a weaker effect $\epsilon_2<\epsilon_1$ applied on both $\beta_s$ and $\beta_w$? This question is equivalent to compare $\gamma^{R_0}_{\beta_w}\epsilon_1 R_0$ to $(\gamma^{R_0}_{\beta_w} + \gamma^{R_0}_{\beta_s})\epsilon_2 R_0$. Note that  
    \begin{equation}\label{eq:r0gbs+gbw}
        \gamma_{\beta_s}^{R_0}+\gamma_{\beta_w}^{R_0}=1.
    \end{equation}
    Direct calculations yield that 
    \begin{equation}
        \gamma^{R_0}_{\beta_w}\epsilon_1 R_0<(\gamma^{R_0}_{\beta_w} + \gamma^{R_0}_{\beta_s})\epsilon_2 R_0 \qquad 
        \Leftrightarrow 
        \qquad 
        \frac{\epsilon_1}{\epsilon_2}< \frac{1}{\gamma^{R_0}_{\beta_w}}
    \end{equation}
    provided (\ref{eq:r0gbs+gbw}). Note that $\epsilon$ must be negative in order to reduce transmission.
    
    Analogous questions can be addressed related to those expressions summing up to $-1$ or $0$.
    
\textbf{Are unexpected management effects possible?}
    We already know that the endemic states bifurcate from the disease-free scenario as $R_0$ crosses the threshold value $1$. The bifurcation can be either subcritical ($R_0<1$ and $\Delta<1$) or supercritical ($R_0 \geq 1$ and $\Delta \geq  1$). The bifurcation direction would make a huge difference, as in the subcritical case $R_0<1$ does not lead necessarily to a disease-free scenario. Changing  any parameter of the model will make vary  simultaneously $R_0$ and $\Delta$ which would result in an unexpected outcome. 
    
    For instance, let us assume that the conditions are such that $R_0 > 1$ and $\Delta > 1$, that is, the system is in the endemic disease  scenario. Assume also that efforts are put into modifying the value of some coefficients to push $R_0$ below $1$. Then, as a result, can $\Delta$ also go below $1$, undergoing a subcritical bifurcation? It would happen if reducing $R_0$ (by any means) would entail a simultaneous reduction in $\Delta$. A necessary condition is that the corresponding sensitivity indices have the same sing. 
     It is apparent that it is not possible if the efforts are put in modifying coefficients $\beta_s$, $\beta_w$, $\gamma_s$ or $\gamma_w$ (see Table \ref{tab:indices}). 
    
    However, it is also apparent that the sign of the sensitivity indices of $R_0$ and $\Delta$ with respect to $a$ (respectively $b$) is the same provided $\beta_w>\beta_s$ (respectively, if $\beta_w<\beta_s$). Indeed direct calculations show that $R_0>1$ and $\Delta>1$ for $N = 70$, $a = 0.27$, $b = 15$, $\beta_S = 3.02$, $\beta_W = 7.2$, $\gamma_S = 20$, $\gamma_W = 6.9$ but  $R_0<1$ and $\Delta<1$ keeping all the previous parameter values but $a=0.17$.\\\newline
    
    We hope the results presented herein would promote further research. On the one hand, we hope experimental scientists find this research interesting and would test the model at their laboratory. On the other hand, more realistic extensions of system (\ref{eq:model}) should  be of interest.

\section*{Declarations:}

\noindent\textbf{Funding:} The work of E. Venturino has been
partially supported by the projects ``Metodi numerici in teoria delle popolazioni'', ``Metodi numerici nelle scienze applicate'' of the
Dipartimento di Matematica ``Giuseppe Peano'' of the
Universit\`a di Torino and the program ``Giner de los R\'ios'' of the Universidad de Alcal\'a. The work of M. Marv\'a has been
partially supported by  Ministerio de Econom\'ia y Competitividad (Spain), project MTM2014-56022-C2-1 and Ministerio de Ciencia, Innovaci\'on y Universidades (Spain), project RTI2018-096884-B-C32-P.\\

\noindent\textbf{Conflicts of interest/Competing interests:} the authors of this work certify  that  they  have  NO  affiliations  with  or  involvement  in  any  organization or entity with any financial interest  or non-financial interest in the subject matter or materials discussed in this manuscript.\\

%
%

\bibliography{main.bib}


\section{Appendix}\label{sec:app}
A first approach to the idea of the  sensitivity of $u$ to $p$ is using the derivative of $u$ with respect to $p$. However, doing so does not allow to fairly compare the sensitivity of $p$ to two different parameters if those parameters are expressed in different units. Defining the sensitivity index as in (\ref{eq:sensit}) fixes this problem (see the most right hand side expression). Furthermore, consider that $u$ depends on the parameters  $p_1,\cdots, p_n$. We may assume without lost of generality that   $p_2,\cdots, p_n$ are held constant, that is equivalent to assume  that $u$ depends only on $p = p_1$. The  Taylor’s expansion approximation of  $u(p)$ to the first order at $p = p_0,$ is given by:

\begin{equation}
    \label{eq:Taylor}
    u(p)\approx u(p_0)+\left.\frac{\partial u(p)}{\partial p}\right|_{p_0}(p-p_0)
\end{equation}
When varying the parameter $p$ by an amount of $\epsilon=(p-p_0)/p_0$, that is to say, from $p_0$ to $p_0+\epsilon p_0$ (\ref{eq:Taylor}) becomes 
\begin{equation}
    \label{eq:Taylor2}
    u(p_0+\epsilon p_0)- u(p_0) \approx \left.\frac{\partial u(p)}{\partial p}\right|_{p_0}\epsilon p_0.
\end{equation}
Taking into account the definition of the sensitivity  index (\ref{eq:sensit}), multiplying and dividing the right hand side of (\ref{eq:Taylor2}) by $u(p_0)$ yields:
\begin{equation}
    \label{eq:Taylor index}
    u(p_0 + \epsilon p_0) - u(p_0) \approx\left.\gamma_p^u\right|_{p_0}\epsilon u(p_0)
\end{equation}
\end{document}